\pdfoutput=1
\documentclass[conference, 10pt]{IEEEtran}
\IEEEoverridecommandlockouts

\usepackage{color}
\usepackage{theorem}
\usepackage{times,amsmath,epsfig}
\usepackage{amssymb}
\usepackage{subcaption}
\usepackage{cite}
\usepackage{url}
\usepackage[capitalise]{cleveref}
\usepackage[ruled,linesnumbered]{algorithm2e}
\usepackage{tikz, pgfplots}
\usepackage{moresize}
\usepackage{array}

\input{mysymbol.sty}

\pgfplotsset{compat=1.17}
\pgfplotstableset{col sep=semicolon}
\usepgfplotslibrary{fillbetween}
\usepgfplotslibrary{groupplots}
\tikzset{every mark/.append style={scale=1.5, solid}, font=\small}
\pgfplotsset{
    width=1.05\textwidth,
    height=5.5cm,
    legend style={
        font=\ssmall ,  
        inner xsep=1pt,
        inner ysep=1pt,
        nodes={inner sep=1pt}},
    legend cell align=left,
    every axis/.append style={line width=.5pt},
 	every axis plot/.append style={line width=1.5pt},
 	every axis y label/.append style={yshift=-4pt}
}

\title{{Non-negative DAG Learning from Time-Series Data}}
\author{{Samuel Rey$^*$
        and~Gonzalo Mateos$^\dagger$} \\
        $^*$Dept. of Signal Theory and Communications,
        Rey Juan Carlos University, Madrid, Spain\\
        Email: samuel.rey.escudero@urjc.es. \\
        $^\dagger$Dept. of Electrical and Computer Engineering, University of Rochester, NY, USA. \\
        Email: gmateosb@ur.rochester.edu
        
\thanks{{This work was supported in part by
the Spanish AEI under Grants PID2019-105032GBI00, TED2021-130347B-I00 and PID2022-136887NBI00 funded by MCIN/AEI/10.13039/501100011033, and by the Community of Madrid via the Ellis Madrid Unit and grants URJC/CAM F1180 and TEC-2024/COM-89.}}}

\begin{document}
%
\maketitle
\begin{abstract}
This work aims to learn the directed acyclic graph (DAG) that captures the instantaneous dependencies underlying a multivariate time series. The observed data follow a linear structural vector autoregressive model (SVARM) with both instantaneous and time-lagged dependencies, where the instantaneous structure is modeled by a DAG to reflect potential causal relationships. While recent continuous relaxation approaches impose acyclicity through smooth constraint functions involving powers of the adjacency matrix, they lead to non-convex optimization problems that are challenging to solve. In contrast, we assume that the underlying DAG has only non-negative edge weights, and leverage this additional structure to impose acyclicity via a convex constraint. This enables us to cast the problem of non-negative DAG recovery from multivariate time-series data as a convex optimization problem in abstract form, which we solve using the method of multipliers. Crucially, the convex formulation guarantees global optimality of the solution. Finally, we assess the performance of the proposed method on synthetic time-series data, where it outperforms existing alternatives.

\end{abstract}

\section{Introduction}\label{s:introduction}
Understanding dependencies in multivariate data requires representations that go beyond simple correlation matrices. 
Directed acyclic graphs (DAGs) have emerged as the canonical representation for modeling complex systems where directionality plays a key role~\cite{marques2020digraphs}, enabling probabilistic reasoning in Bayesian networks and causal inference tasks~\cite{spirtes2001causation,koller2009probabilistic,peters2017elements,yao2021survey}. 
Consequently, DAGs are ubiquitous in applications requiring the identification of driving forces, ranging from machine learning~\cite{koller2009probabilistic,yu2019dag,rey2024convolutional} and signal processing~\cite{seifert2023causal,misiakos2024learning}, to biology~\cite{lucas2004bayesian} and genetics~\cite{zhang2013integrated}. 
Despite their widespread adoption, the underlying DAG structure is rarely known \emph{a priori}. 
This necessitates the development of structure learning methods capable of recovering the graph from observational data.

Inferring the graph topology from nodal observations constitutes a central problem, connecting the statistical properties of data to the underlying graph structure~\cite{mateos2019connecting,giannakis18}. 
For undirected graphs, noteworthy approaches include smoothness priors~\cite{kalofolias2016learn,dong2016learning,saboksayr2021accelerated}, Gaussian graphical models~\cite{friedman2008sparse,egilmez2017graph,rey2023enhanced,rey2025online}, and graph stationary models~\cite{segarra2017network,shafipour2020online,roddenberry2021network,buciulea2022learning,navarro2024joint}, among others. 
However, when the focus shifts to DAGs, structural equation models (SEMs) become the prevailing formalism~\cite{zheng2018dags,wei2020dags,bello2022dagma,cheng2025isit,saboksayr2023colide}. 
Strictly enforcing the acyclicity constraint renders DAG structure learning a combinatorial, NP-hard endeavor~\cite{maxwell1997efficient,chickering2004large}. 
This complexity is further exacerbated in time-series settings, as temporal lags drastically increase the dimensionality of the search space, rendering traditional discrete search methods computationally prohibitive~\cite{pamfil2020dynotears,murphy2002dynamic}.

Recent advancements have revolutionized the field of DAG structure learning, with impact to causal discovery.
The seminal work in~\cite{zheng2018dags} reformulated the combinatorial problem of learning the DAG structure from nodal observations into a continuous optimization task, enabling efficient exploration of the DAG space.
This breakthrough sparked a wave of variants exploring alternative acyclicity constraints and score functions~\cite{wei2020dags,ng2020role,bello2022dagma,saboksayr2023colide}.
In the time-series domain, dynamic Bayesian networks (DBNs) have long served as the standard for modeling temporal processes, particularly over discrete state spaces~\cite{dean1989model,murphy2002dynamic}.
Building on structural vector autoregressive models (SVARMs)~\cite{swanson1997impulse,giannakis18}, recent approaches like DYNOTEARS~\cite{pamfil2020dynotears,gao2022idyno} extended the continuous acyclicity constraint to the dynamic setting.
Further extensions incorporating sparsity-promoting score functions have also been proposed~\cite{misiakos2025learning}.
However, a fundamental limitation persists: the resulting optimization landscape is inherently non-convex.
Consequently, state-of-the-art methods often rely on heuristics or specific initializations~\cite{bello2022dagma}, frequently converging to local minima rather than the global optimum~\cite{deng2023global}.

In this work, we surmount these optimization challenges by harnessing recent advances in non-negative DAG structure learning~\cite{rey2025non}. 
Specifically, we tackle the problem of recovering DAGs with non-negative edge weights from time-series data governed by linear SVARMs. Crucially, the non-negativity constraint enables characterizing acyclicity via a \emph{convex} function. 
We formulate the inference task as an \emph{abstract convex optimization problem} and propose an iterative algorithm based on the method of multipliers to solve it. 
Unlike previous approaches, our formulation warrants the recovery of the global minimizer, eliminating the sensitivity to initialization. 
The merits of this approach are demonstrated in our numerical evaluation, where the proposed method outperforms state-of-the-art alternatives on synthetic datasets.

\section{DAG structure learning from time series}\label{s:fundamentals}
Let $\ccalD = (\ccalV, \ccalE)$ denote a directed graph with a set of $N$ nodes $\ccalV$ and edges $\ccalE \subseteq \ccalV \times \ccalV$.
The graph $\ccalD$ is a DAG if it contains no directed cycles.
The connectivity of the DAG is captured by the weighted adjacency matrix $\bbW \in \reals^{N \times N}$, where $W_{ij} \neq 0$ if and only if $(i, j) \in \ccalE$.
Hence, the entry $W_{ij}$ denotes the weight of the directed link from node $i$ to node $j$.
Finally, we define a graph signal on $\ccalD$ as a vector $\bbx \in \reals^N$, where the $i$-th entry $x_i$ denotes the signal value at node $i$.

We consider the problem of learning the causal structure from a time series of graph signals $\{\bbx_t\}_{t=1}^T$, where $\bbx_t \in \reals^N$ represents the graph signal at time index $t$.
We assume the data generation process adheres to a SVARM of order $P$~\cite{swanson1997impulse}. 
Formally, for each time instant $t$, the signal $\bbx_t$ is modeled as
\begin{equation} \label{eq:svarm_vector}
	\bbx_t = \bbW^\top \bbx_t + \sum_{p=1}^P \bbA_p^\top \bbx_{t-p} + \bbz_t,
\end{equation}
where the matrix $\bbW$ encodes the instantaneous dependencies, and the set of matrices $\{\bbA_p\}_{p=1}^P$, with $\bbA_p \in \reals^{N \times N}$, captures the time-lagged influences from the $p$-th previous step.
The term $\bbz_t$ represents a vector of exogenous noises, assumed to be an i.i.d. random vector with covariance matrix $\bbSigma_\bbz = \sigma^2\bbI$.

We can derive a compact matrix formulation for the SVARM defined in~\eqref{eq:svarm_vector}.
To that end, we collect the observed signals in the matrix $\bbX := [\bbx_1, \ldots, \bbx_T] \in \reals^{N \times T}$.
Similarly, let $\bbY_p \in \reals^{N \times T}$ denote the time-lagged data matrix for lag $p$, constructed by shifting the columns of $\bbX$ appropriately.
By stacking the lagged matrices vertically as $\bbY := [\bbY_1^\top, \ldots, \bbY_P^\top]^\top \in \reals^{NP \times T}$ and the corresponding coefficients as $\bbA := [\bbA_1^\top, \ldots, \bbA_P^\top]^\top \in \reals^{NP \times N}$, the signal model takes the form
\begin{equation} \label{eq:svarm_matrix}
	\bbX = \bbW^\top \bbX + \bbA^\top \bbY + \bbZ,
\end{equation}
where $\bbZ \in \reals^{N \times T}$ collects the noise vectors.
Note that the standard linear SEM is recovered as a particular case of~\eqref{eq:svarm_matrix} when the autoregressive order is $P=0$ (i.e., $\bbA \equiv \mathbf{0}$).
As stated in~\cite{pamfil2020dynotears}, the SVARM is identifiable under relatively mild assumptions.
In particular, $\bbW$ is identifiable whenever the exogenous noise vectors $\bbz_t$ are either non-Gaussian or standard Gaussian $\bbz_t \sim \ccalN(\bbzero,\sigma^2\bbI)$.
Identifiability of $\bbA$ follows from standard results on vector autoregressive models.

With the previous definitions in place, the joint inference of the instantaneous DAG structure $\bbW$ and the time-lagged dependencies $\bbA$ from the observed time series $\bbX$ can be cast as the following optimization problem
\begin{equation}\label{eq:dag_learning} 
    \min_{\bbW, \bbA} \quad F \left( \bbW, \bbA; \bbX \right) \quad \text{subject to} \quad \bbW \in \mbD,
\end{equation}
where $F(\cdot)$ denotes a data-dependent score function quantifying the goodness of fit, and $\mbD$ represents the set of adjacency matrices corresponding to a DAG.
For a valid causal interpretation in the absence of latent variables, $\bbW$ must be a DAG. 
Conversely, the lagged coefficient matrices collected in $\bbA$ are not required to be acyclic, as they encapsulate how past system states influence the current state at time $t$.

The optimization problem in \eqref{eq:dag_learning} is challenging to solve due to the non-convex and combinatorial nature of the constraint $\bbW \in \mbD$.
Recent advances propose addressing this issue by replacing the discrete set $\mbD$ with a continuous \emph{acyclicity constraint} of the form $h(\bbW) = 0$, where $h: \reals^{N \times N} \mapsto \reals$ is a differentiable function whose zero level set coincides with the space of DAGs $\mbD$.
This approach was pioneered in~\cite{zheng2018dags} and subsequently followed by several works proposing alternative continuous acyclicity functions~\cite{bello2022dagma,wei2020dags}.
Replacing the combinatorial constraint with $h(\bbW) = 0$ constitutes a significant advancement, enabling the use of standard continuous optimization methods.
However, existing acyclicity functions $h(\bbW)$ are generally non-convex. 
Consequently, the resulting optimization landscape remains fraught with local minima, hindering the recovery of the true DAG structure.
To overcome this limitation, we henceforth assume $\bbW$ has \emph{non-negative weights}.
This allows us to leverage a convex acyclicity function and develop an algorithm to obtain the \emph{global minimizer} of an optimization problem equivalent to~\eqref{eq:dag_learning}.

\section{Learning non-negative DAGs from SVAR data}
We approach the DAG structure learning problem by enforcing a non-negativity constraint on the entries of $\bbW$.
This assumption is practically relevant, extending to binary DAGs and other pragmatic settings~\cite{seifert2023causal}.
Crucially, this additional structure simplifies the optimization landscape of~\eqref{eq:dag_learning}, enabling the use of a convex acyclicity function to recover the global minimizer of the score function.
In the sequel, we detail the proposed optimization problem and the algorithmic solution.

Central to modern DAG learning methods is the use of smooth constraints to enforce the absence of cycles.
As discussed in~\cite{zheng2018dags}, an effective acyclicity function $h(\bbW)$ must be differentiable, computationally efficient, and satisfy $h(\bbW) = 0$ if and only if $\bbW \in \mathbb{D}$.
In this work, we leverage recent advancements introduced in~\cite{rey2025non}, which demonstrate that imposing non-negativity on $\bbW$ allows for a \emph{convex} characterization of acyclicity.
The details about this acyclicity function are provided in the following proposition.
\begin{proposition}[\cite{rey2025non}]\label{prop:cvx_logdet}
	Let $\bbW \in \reals_+^{N \times N}$ be a non-negative matrix with spectral radius bounded by $\rho(\bbW) < s$ for some $s > 0$. 
	Consider the function
	\begin{equation}\label{eq:cvx_logdet}
		h(\bbW) := N\log(s) - \log\det(s\bbI - \bbW).
	\end{equation}
	Then, $h(\bbW)$ is convex over its domain, differentiable with gradient $\nabla h (\bbW) = \left(s\bbI - \bbW \right)^{-\top}$, and satisfies $h(\bbW) \geq 0$.
	Furthermore, $h(\bbW) = 0$ if and only if $\bbW \in \mathbb{D}$.
\end{proposition}

When $\bbX$ adheres to a SVARM, a standard choice for the score function is the least-squares data-fidelity term augmented with $\ell_1$-norm regularization to promote sparsity on both $\bbW$ and $\bbA$. Leveraging this convex score function, the continuous acyclicity constraint, and enforcing entry-wise non-negativity on both $\bbW$ and $\bbA$, the DAG structure learning problem can be alternatively formulated as
\begin{alignat}{3}
		\{\hbW, \hbA\} = \quad & \arg\min_{{\bbW}, {\bbA}} \quad
		&& \frac{1}{2T} \| \bbX - \bbW^\top\bbX - \bbA^\top\bbY \|_F^2 \nonumber \\
        & && + \lambda_{\bbW} \sum_{i,j} W_{ij} + \lambda_{\bbA} \sum_{k,l} A_{kl}\nonumber \\
		& \text{subject to} && \bbW \geq 0, \;\; \bbA \geq 0, \;\;h(\bbW) = 0. \label{eq:opt_problem}
\end{alignat}
Here, the regularization parameters $\lambda_{\bbW}, \lambda_\bbA > 0$ control the trade-off between data fidelity and sparsity for the instantaneous and time-lagged coefficients, respectively.
The term $h(\bbW)$ denotes the convex acyclicity function defined in~\eqref{eq:cvx_logdet}.
Note that, due to the non-negativity constraints, the standard $\ell_1$ penalties $\| \bbW \|_1$ and $\| \bbA \|_1$ simplify to the linear sums of the entries appearing in the objective function.
It is worth noting that while we assume both $\bbW$ and $\bbA$ to be non-negative for consistency, only the non-negativity of $\bbW$ is required for the convexity of the acyclicity function.
Therefore, the proposed framework can readily accommodate signed lagged coefficients by simply removing the constraint $\bbA \geq 0$ and replacing the linear sum penalty with the standard $\ell_1$-norm.

Ensuring the acyclicity of $\bbW$ via a convex function such as $h$ confers significant advantages. 
Foremost, the convexity of $h$ renders~\eqref{eq:opt_problem} an \emph{abstract convex optimization problem}~\cite{boyd2004convex}, thereby guaranteeing the recovery of the global minimizer.
To see this, recall from Proposition~\ref{prop:cvx_logdet} that $h(\bbW) \geq 0$, thus the equality constraint $h(\bbW) = 0$ is mathematically equivalent to the inequality constraint $h(\bbW) \leq 0$.
Since the sublevel sets of a convex function form a convex set, the feasible region of~\eqref{eq:opt_problem} is convex.
In turn, the ability to recover the global minimum, combined with the identifiability of the SVARM in~\eqref{eq:svarm_matrix} under mild assumptions, paves the way for establishing strong statistical guarantees for the estimated DAG, a promising avenue for future research.

\subsection{DAG structure learning via the method of multipliers}\label{sec:algorithm}

Estimating the instantaneous and time-lagged dependencies encoded in $\bbW$ and $\bbA$ requires solving the constrained optimization problem introduced in \eqref{eq:opt_problem}.
To that end, we rely on the method of multipliers~\cite[Ch. 4.2]{bertsekas1997dynamic}, an iterative approach based on the augmented Lagrangian to tackle equality-constrained problems, which offers convergence guarantees.

The augmented Lagrangian of~\eqref{eq:opt_problem} is given by
\begin{align} \label{eq:aug_lagrangian}
	L_c (\bbW, &\bbA, \alpha) = \frac{1}{2T}\| \bbX - \bbW^\top\bbX -\bbA^\top\bbY \|_F^2 +\lambda_{\bbW} \sum_{i,j} W_{ij} \nonumber \\
    &\quad +\lambda_{\bbA} \sum_{k,l} A_{kl} + \alpha h(\bbW) + \frac{c}{2} h_{ldet}^2(\bbW),
\end{align}
where $\alpha \in \reals_+$ is the Lagrange multiplier associated with the acyclicity constraint, and $c \in\reals_+$ serves as the quadratic penalty parameter.
Observe that the augmented Lagrangian $L_c$ is convex with respect to the primal variables $\bbW$ and $\bbA$.
This follows from Proposition~\ref{prop:cvx_logdet}, which guarantees that $h(\bbW) \geq 0$, and hence, the term $h^2(\bbW)$ represents the composition of a convex function with a convex, non-decreasing function~\cite{boyd2004convex}.
Finally, note that the non-negativity constraints $\bbW \geq 0$ and $\bbA \geq 0$ are not explicitly included in $L_c$ since they can be enforced through a simple projection.

Given the augmented Lagrangian formulation, to solve~\eqref{eq:opt_problem} we adopt an iterative approach where we perform the following sequence of steps for $k=1,\ldots,K$ iterations.

\vspace{2mm}
\noindent
\textbf{Step 1.}
We update the primal variables $\bbW^{(k+1)}$ and $\bbA^{(k+1)}$ by minimizing
\begin{equation}\label{eq:step1}
	\left\{\bbW^{(k+1)}, \bbA^{(k+1)} \right\}= \arg\min_{\bbW \geq 0, \bbA \geq 0} L_{c^{(k)}} (\bbW, \bbA, \alpha^{(k)}).
\end{equation}
Due to the convexity of $L_{c^{(k)}}$, we can recover the global minimizer using standard convex optimization methods, such as (accelerated) projected gradient descent.

\vspace{2mm}
\noindent
\textbf{Step 2.}
The update of the Lagrange multiplier $\alpha^{(k+1)}$ depends on the current constraint violation, given by
\begin{equation}\label{eq:step2}
	\alpha^{(k+1)} = \alpha^{(k)} + c^{(k)}h(\bbW^{(k+1)}).
\end{equation}
This update is equivalent to a gradient ascent step, as the constraint violation corresponds to the gradient of $L_{c^{(k)}}$ with respect to $\alpha$.

\vspace{2mm}
\noindent
\textbf{Step 3.}
The penalty parameter $c^{(k)}$ is progressively increased to ensure that the constraint $h(\bbW) = 0$ is satisfied as $K \to \infty$.
A typical update scheme is given by
\begin{equation}\label{eq:step3}
	c^{(k+1)} = 
	\left\{ \begin{array}{lc} 
		\beta c^{(k)} & \mathrm{if} \;\; h(\bbW^{(k+1)}) > \gamma h(\bbW^{(k)}) \\
		c^{(k)} & \mathrm{otherwise},
	 \end{array} \right.
\end{equation}
where $0 <\gamma < 1$ and $\beta > 1$ are user-defined constants.
Intuitively, $c^{(k)}$ is increased only if the constraint violation is not decreased by a factor of $\gamma$.

Upon completion of the sequence of iterations, the estimated DAG structure and time-lagged dependencies are given by $\hbW = \bbW^{(K)}$ and $\hbA = \bbA^{(K)}$, respectively.
The convexity of the augmented Lagrangian guarantees that the tuple $\{\bbW^{(k)}, \bbA^{(k)}\}$ corresponds to the global minimum of the subproblem in \eqref{eq:step1} at each iteration $k$.
Therefore, as established in~\cite[Prop. 4.2.1]{bertsekas1997dynamic}, every limit point of the sequence $\{\bbW^{(k)}, \bbA^{(k)}\}$ constitutes a global minimum of the constrained problem in \eqref{eq:opt_problem}.

While the method of multipliers was also employed in~\cite{zheng2018dags}, subsequent work identified a significant drawback in their formulation~\cite{wei2020dags}.
Specifically, every DAG constitutes a stationary point of the acyclicity constraint used in~\cite{zheng2018dags}, implying that $\nabla h_{\text{notears}} (\bbW) = \mathbf{0}$ for any $\bbW \in \mathbb{D}$.
This implies that a feasible solution can satisfy the KKT optimality conditions only if it is also a stationary point of the unconstrained objective, posing algorithmic challenges for finding the true constrained minimizer.
In contrast, observe that DAGs are not stationary points of our acyclicity function (i.e., $\nabla h (\bbW) \neq \mathbf{0}$).
Consequently, our method avoids these algorithmic pitfalls.


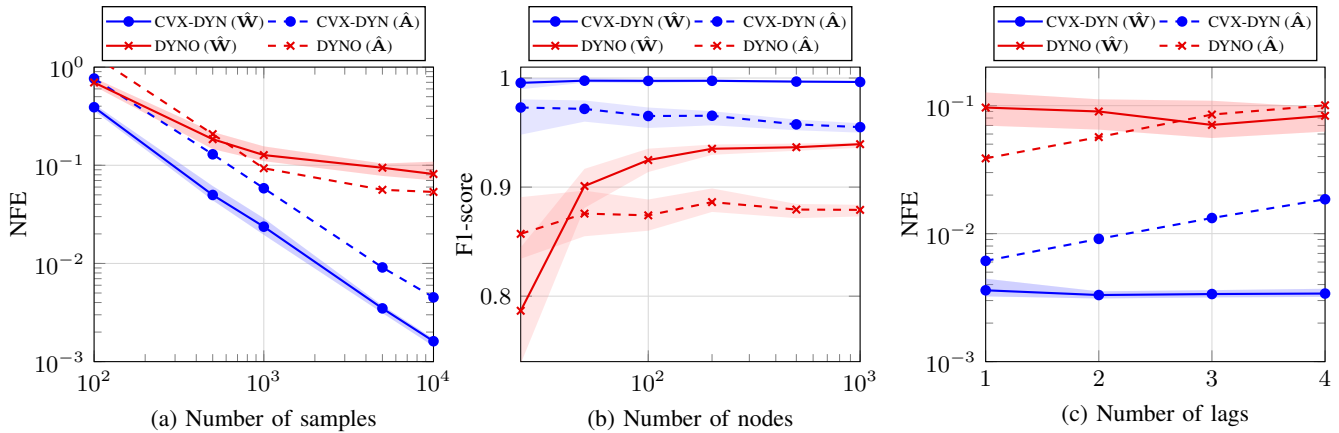
\begin{figure*}[!t]
	\centering
	\begin{subfigure}[t]{0.32\textwidth}
		\centering
		\begin{tikzpicture}[baseline,scale=1]
    \pgfplotstableread{data/samples/samples_errW_med.csv}\errWmed
    \pgfplotstableread{data/samples/samples_errW_prctile25.csv}\prctWlow
    \pgfplotstableread{data/samples/samples_errW_prctile75.csv}\prctWhigh
    
    \pgfplotstableread{data/samples/samples_errA_med.csv}\errAmed
    \pgfplotstableread{data/samples/samples_errA_prctile25.csv}\prctAlow
    \pgfplotstableread{data/samples/samples_errA_prctile75.csv}\prctAhigh
    
    \pgfmathsetmacro{\opacity}{0.2} 

    \begin{loglogaxis}[
        xlabel={(a) Number of samples},
        ylabel={NFE},
        xmin=100,  
        xmax=10000,
        xtick = {100, 1000, 10000},
        ymin = 1e-3, 
        ymax = 1,
        grid style={gray!30},
        grid=major,
        legend style={
            at={(.5,1.02)},
            anchor=south},
        legend columns=2,
        ]

        
        \addplot [name path=adamWtop, draw=none, forget plot] table [x=xaxis, y=Adam] \prctWhigh;
        \addplot [name path=adamWbot, draw=none, forget plot] table [x=xaxis, y=Adam] \prctWlow;
        \addplot [blue!50, opacity=\opacity, forget plot] fill between[of=adamWtop and adamWbot];
        \addplot[blue, thick, mark=*, mark options={scale=0.8}] table [x=xaxis, y=Adam] {\errWmed};
        \addlegendentry{CVX-DYN ($\hbW$)}

        \addplot [name path=adamAtop, draw=none, forget plot] table [x=xaxis, y=Adam] \prctAhigh;
        \addplot [name path=adamAbot, draw=none, forget plot] table [x=xaxis, y=Adam] \prctAlow;
        \addplot [blue!50, opacity=\opacity, forget plot] fill between[of=adamAtop and adamAbot];
        \addplot[blue, thick, dashed, mark=*, mark options={solid, scale=0.8}] table [x=xaxis, y=Adam] {\errAmed};
        \addlegendentry{CVX-DYN ($\hbA$)}


        \addplot [name path=dynWtop, draw=none, forget plot] table [x=xaxis, y=DYNOTEARS] \prctWhigh;
        \addplot [name path=dynWbot, draw=none, forget plot] table [x=xaxis, y=DYNOTEARS] \prctWlow;
        \addplot [red!50, opacity=\opacity, forget plot] fill between[of=dynWtop and dynWbot];
        \addplot[red!90!black, thick, mark=x, mark options={scale=1}] table [x=xaxis, y=DYNOTEARS] {\errWmed};
        \addlegendentry{DYNO ($\hbW$)}

        \addplot [name path=dynAtop, draw=none, forget plot] table [x=xaxis, y=DYNOTEARS] \prctAhigh;
        \addplot [name path=dynAbot, draw=none, forget plot] table [x=xaxis, y=DYNOTEARS] \prctAlow;
        \addplot [red!50, opacity=\opacity, forget plot] fill between[of=dynAtop and dynAbot];
        \addplot[red!90!black, thick, dashed, mark=x, mark options={solid, scale=1}] table [x=xaxis, y=DYNOTEARS] {\errAmed};
        \addlegendentry{DYNO ($\hbA$)}

    \end{loglogaxis}
\end{tikzpicture}	
	\end{subfigure}
	\begin{subfigure}[t]{0.32\textwidth}
		\centering
		\begin{tikzpicture}[baseline,scale=1]
    \pgfplotstableread{data/nodes/nodes_fsW_med.csv}\errWmed
    \pgfplotstableread{data/nodes/nodes_fsW_prctile25.csv}\prctWlow
    \pgfplotstableread{data/nodes/nodes_fsW_prctile75.csv}\prctWhigh
    
    \pgfplotstableread{data/nodes/nodes_fsA_med.csv}\errAmed
    \pgfplotstableread{data/nodes/nodes_fsA_prctile25.csv}\prctAlow
    \pgfplotstableread{data/nodes/nodes_fsA_prctile75.csv}\prctAhigh
    
    \pgfmathsetmacro{\opacity}{0.2} 

    \begin{semilogxaxis}[
        xlabel={(b) Number of nodes},
        ylabel={F1-score},
        xmin=25,  
        xmax=1000,
        xtick = {100, 1000, 10000},
        ymin = 0.74, 
        ymax = 1.01,
        grid style={gray!30},
        grid=major,
        legend style={
            at={(.5,1.02)},
            anchor=south},
        legend columns=2,
        ]

        
        \addplot [name path=adamWtop, draw=none, forget plot] table [x=xaxis, y=Adam] \prctWhigh;
        \addplot [name path=adamWbot, draw=none, forget plot] table [x=xaxis, y=Adam] \prctWlow;
        \addplot [blue!50, opacity=\opacity, forget plot] fill between[of=adamWtop and adamWbot];
        \addplot[blue, thick, mark=*, mark options={scale=0.8}] table [x=xaxis, y=Adam] {\errWmed};
        \addlegendentry{CVX-DYN ($\hbW$)}

        \addplot [name path=adamAtop, draw=none, forget plot] table [x=xaxis, y=Adam] \prctAhigh;
        \addplot [name path=adamAbot, draw=none, forget plot] table [x=xaxis, y=Adam] \prctAlow;
        \addplot [blue!50, opacity=\opacity, forget plot] fill between[of=adamAtop and adamAbot];
        \addplot[blue, thick, dashed, mark=*, mark options={solid, scale=0.8}] table [x=xaxis, y=Adam] {\errAmed};
        \addlegendentry{CVX-DYN ($\hbA$)}


        \addplot [name path=dynWtop, draw=none, forget plot] table [x=xaxis, y=DYNOTEARS] \prctWhigh;
        \addplot [name path=dynWbot, draw=none, forget plot] table [x=xaxis, y=DYNOTEARS] \prctWlow;
        \addplot [red!50, opacity=\opacity, forget plot] fill between[of=dynWtop and dynWbot];
        \addplot[red!90!black, thick, mark=x, mark options={scale=1}] table [x=xaxis, y=DYNOTEARS] {\errWmed};
        \addlegendentry{DYNO ($\hbW$)}

        \addplot [name path=dynAtop, draw=none, forget plot] table [x=xaxis, y=DYNOTEARS] \prctAhigh;
        \addplot [name path=dynAbot, draw=none, forget plot] table [x=xaxis, y=DYNOTEARS] \prctAlow;
        \addplot [red!50, opacity=\opacity, forget plot] fill between[of=dynAtop and dynAbot];
        \addplot[red!90!black, thick, dashed, mark=x, mark options={solid, scale=1}] table [x=xaxis, y=DYNOTEARS] {\errAmed};
        \addlegendentry{DYNO ($\hbA$)}

    \end{semilogxaxis}
\end{tikzpicture}
	\end{subfigure}
	\begin{subfigure}[t]{0.32\textwidth}
		\centering
		\begin{tikzpicture}[baseline,scale=1]
    \pgfplotstableread{data/lags/lags_errW_med.csv}\errWmed
    \pgfplotstableread{data/lags/lags_errW_prctile25.csv}\prctWlow
    \pgfplotstableread{data/lags/lags_errW_prctile75.csv}\prctWhigh
    
    \pgfplotstableread{data/lags/lags_errA_med.csv}\errAmed
    \pgfplotstableread{data/lags/lags_errA_prctile25.csv}\prctAlow
    \pgfplotstableread{data/lags/lags_errA_prctile75.csv}\prctAhigh
    
    \pgfmathsetmacro{\opacity}{0.2} 

    \begin{semilogyaxis}[
        xlabel={(c) Number of lags},
        ylabel={NFE},
        xmin=1,  
        xmax=4,
        ymin = 1e-3, 
        ymax = .2,
        grid style={gray!30},
        grid=major,
        legend style={
            at={(.5,1.02)},
            anchor=south},
        legend columns=2,
        ]

        
        \addplot [name path=adamWtop, draw=none, forget plot] table [x=xaxis, y=Adam] \prctWhigh;
        \addplot [name path=adamWbot, draw=none, forget plot] table [x=xaxis, y=Adam] \prctWlow;
        \addplot [blue!50, opacity=\opacity, forget plot] fill between[of=adamWtop and adamWbot];
        \addplot[blue, thick, mark=*, mark options={scale=0.8}] table [x=xaxis, y=Adam] {\errWmed};
        \addlegendentry{CVX-DYN ($\hbW$)}

        \addplot [name path=adamAtop, draw=none, forget plot] table [x=xaxis, y=Adam] \prctAhigh;
        \addplot [name path=adamAbot, draw=none, forget plot] table [x=xaxis, y=Adam] \prctAlow;
        \addplot [blue!50, opacity=\opacity, forget plot] fill between[of=adamAtop and adamAbot];
        \addplot[blue, thick, dashed, mark=*, mark options={solid, scale=0.8}] table [x=xaxis, y=Adam] {\errAmed};
        \addlegendentry{CVX-DYN ($\hbA$)}


        \addplot [name path=dynWtop, draw=none, forget plot] table [x=xaxis, y=DYNOTEARS] \prctWhigh;
        \addplot [name path=dynWbot, draw=none, forget plot] table [x=xaxis, y=DYNOTEARS] \prctWlow;
        \addplot [red!50, opacity=\opacity, forget plot] fill between[of=dynWtop and dynWbot];
        \addplot[red!90!black, thick, mark=x, mark options={scale=1}] table [x=xaxis, y=DYNOTEARS] {\errWmed};
        \addlegendentry{DYNO ($\hbW$)}

        \addplot [name path=dynAtop, draw=none, forget plot] table [x=xaxis, y=DYNOTEARS] \prctAhigh;
        \addplot [name path=dynAbot, draw=none, forget plot] table [x=xaxis, y=DYNOTEARS] \prctAlow;
        \addplot [red!50, opacity=\opacity, forget plot] fill between[of=dynAtop and dynAbot];
        \addplot[red!90!black, thick, dashed, mark=x, mark options={solid, scale=1}] table [x=xaxis, y=DYNOTEARS] {\errAmed};
        \addlegendentry{DYNO ($\hbA$)}

    \end{semilogyaxis}
\end{tikzpicture}
	\end{subfigure}
		\vspace{-0.15cm}
	\caption{Performance evaluation of the proposed method (CVX-DYN) compared to DYNOTEARS (we use DYNO in the legends for clarity) across different scenarios. (a) NFE of the estimated weights $\hbW$ and $\hbA$ as a function of the time-series length $T$. (b) F1-score of $\hbW$ and $\hbA$ as the number of nodes $N$ increases. (c) NFE for varying autoregressive orders $P$. 
    }
    \vspace{-.4cm}
    \label{fig:exps}
\end{figure*}

\section{Numerical experiments}
Here, we evaluate the performance of the proposed method across various scenarios.
The code for the algorithm and all implementation details is publicly available on GitHub\footnote{\url{https://github.com/reysam93/cvx_dyn_dag}}.

We assess the estimation accuracy using two metrics.
We measure the normalized Frobenius error (NFE) calculated as
\begin{equation}
    \text{NFE}(\hbW, \bbW^*) = \frac{\| \bbW^* - \hbW \|_F^2}{\| \bbW^* \|_F^2},
\end{equation}
as well as the F1-score, defined as the harmonic mean of precision and recall computed on the support of the adjacency matrices, to evaluate structure recovery.
We report these metrics for both the instantaneous effects $\hbW$ and the lagged effects $\hbA$.
In all experiments, we report the median values along with the 25th and 75th percentiles obtained from 50 independent realizations.
We compare our approach with DYNOTEARS~\cite{pamfil2020dynotears}, a closely related alternative that also assumes a SVARM and estimates $\bbW$ and $\bbA$ simultaneously, but leverages a continuous non-convex acyclicity constraint.

Unless otherwise stated, we simulate Erd\H{o}s-Rényi (ER) DAGs with $N=50$ nodes and an average degree of 4, generated by sampling lower-triangular adjacency matrices and randomly permuting their rows and columns.
The $P=2$ time-lagged matrices $\{\bbA_p\}_{p=1}^P$ are sampled as ER graphs with an average degree of 1.
Edge weights for $\bbW$ and $\bbA$ are sampled uniformly from the interval $[0.1, 0.5]$.
Following~\cite{pamfil2020dynotears},we also apply an exponential decay $e^{-1.5p}$  to $\bbA_p$ to promote the stability of the SVARM process.
Finally, time series data is generated according to \eqref{eq:svarm_matrix} with length $T=5000$, driven by i.i.d. noise $\bbZ$ sampled from a standard Gaussian distribution.

\vspace{2mm}
\noindent
\textbf{Test case 1 - Number of samples.}
The first experiment evaluates the estimation accuracy of the proposed method (CVX-DYN) relative to DYNOTEARS as the sample size $T$ increases.
As illustrated in \cref{fig:exps} (a), leveraging a convex acyclicity constraint consistently yields superior performance compared to the baseline.
Notably, while the error for DYNOTEARS saturates, the error associated with our method exhibits a continuous decay, converging towards zero as the number of samples increases. This empirical behavior supports the intuition that, since the SVARM is identifiable and our convex formulation recovers the global minimum, the method should be capable of recovering the true underlying structure given sufficient data.
Additionally, we observe that the error for the lagged coefficients $\hbA$ is consistently higher than for the instantaneous weights $\hbW$.
This discrepancy can be attributed to the data generation process, where the exponential decay applied to $\bbA_p$ reduces the influence of lagged dependencies, making them inherently more challenging to estimate.

\vspace{2mm}
\noindent
\textbf{Test case 2 - Number of nodes.}
Next, we test the scalability of the methods by analyzing the F1-score for $\hbW$ and $\hbA$ as the network size $N$ increases (see \cref{fig:exps} (b)).
The results demonstrate that CVX-DYN consistently outperforms the baseline, further validating the advantages of our convex approach.
Remarkably, our method achieves a perfect F1-score of 1 for the instantaneous matrix $\hbW$, indicating robust support recovery even in high-dimensional regimes where the number of nodes ($N=1000$) becomes comparable to the sample size ($T=5000$).
Regarding the lagged dependencies $\hbA$, the performance follows the trend observed in the previous test case.
While our method still outperforms the baseline, the recovery is inherently more challenging due to the weaker influence of the lagged components.

\vspace{2mm}
\noindent
\textbf{Test case 3 - Time-lagged dependencies.}
Finally, in \cref{fig:exps}~(c) we assess how the autoregressive order $P$ impacts the performance of the DAG learning alternatives.
The results illustrate that the error of the estimated $\hbW$ remains stable for both methods (CVX-DYN and DYNOTEARS), irrespective of the value of $P$, which is likely partially due to the exponential decay strategy followed to generate the data. 
In contrast, we observe that the error for $\hbA$ increases monotonically with the number of lags, underscoring that accounting for longer temporal dependencies results in a more challenging inference problem.
In addition, \cref{fig:exps} (c) demonstrates that CVX-DYN attains a consistently low estimation error for both $\hbW$ and $\hbA$, which is approximately one order of magnitude smaller than the error attained by DYNOTEARS, emphasizing the merits of our proposed convex approach.

\section{Conclusions}
This paper addressed the prominent task of jointly learning the DAG structure, which models instantaneous dependencies, and the matrix $\bbA$, which encodes time-lagged dependencies, from nodal observations governed by a SVARM. By focusing on the estimation of a non-negative matrix $\hbW$, we formulated the DAG learning task as a continuous optimization problem in a convex abstract form.
Harnessing the specific structure of our formulation, we proposed an iterative algorithm based on the method of multipliers. To the best of our knowledge, this is the first method for learning DAGs from time series that guarantees the recovery of the global minimizer. We argue that our approach avoids algorithmic issues present in competing DAG learning schemes also based on augmented Lagrangian methods. Finally, we corroborate the algorithmic advantages of our proposal through reproducible numerical experiments on synthetic data, where it consistently outperforms state-of-the-art methods.

\bibliographystyle{IEEEbib}
\bibliography{myIEEEabrv,biblio}

\end{document}